\documentclass[preprint,5p,times]{elsarticle}
\usepackage{lineno}
\usepackage[T1]{fontenc}
\usepackage[utf8]{inputenc}
\usepackage{graphicx}
\usepackage{amsmath}
\usepackage{textcomp}
\journal{NIM A}

\begin{document}
\begin{frontmatter}
\title{Large Scales -- Long Times: Adding High Energy Resolution to SANS}
\author[frm,e21]{G. Brandl}
\author[frm,e21]{R. Georgii}
\ead{Corresponding author: Robert.Georgii@frm2.tum.de}
\author[frm,e21]{W. Häußler}
\author[ETH]{S. Mühlbauer}
\author[e21]{P. Böni}
\address[frm]{Forschungsneutronenquelle Heinz Maier-Leibnitz,
  Technische Universität München, Lichtenbergstr.\ 1, 85747 Garching, Germany}
\address[e21]{Physik Department E21,
  Technische Universität München, James-Franck-Str., 85747 Garching, Germany}
\address[ETH]{Neutron Scattering and Magnetism Group, Laboratorium für
  Festkörperphysik, ETH Zürich, 8093 Zürich, Switzerland}
\date{\today}
\begin{abstract}
  The Neutron Spin Echo (NSE) variant MIEZE (Modulation of IntEnsity by Zero
  Effort), where all beam manipulations are performed before the sample
  position, offers the possibility to perform low background SANS measurements
  in strong magnetic fields and depolarising samples.  However, MIEZE is
  sensitive to differences $\Delta L$ in the length of neutron flight paths
  through the instrument and the sample.  In this article, we discuss the major
  influence of $\Delta L$ on contrast reduction of MIEZE measurements and its
  minimisation.  Finally we present a design case for enhancing a small-angle
  neutron scattering (SANS) instrument at the planned European Spallation Source
  (ESS) in Lund, Sweden, using a combination of MIEZE and other TOF options,
  such as TISANE offering time windows from ns to minutes.  The proposed
  instrument would allow obtaining an excellent energy- and $Q$-resolution
  straightforward to \textmu s for 0.01\,\AA$^{-1}$, even in magnetic fields,
  depolarising samples as they occur in soft matter and magnetism while keeping
  the instrumental effort and costs low.

\end{abstract}
\end{frontmatter}

\section{Introduction}
The Neutron Spin Echo technique (NSE) \cite{Mezei:72} in its different variants
is a unique method for measuring dynamic processes in soft matter
\cite{Richter:05} and spin excitations in magnetic systems \cite{Mezei:84}.  As
it allows for the decoupling of the incoming wavelength distribution and the
energy resolution, typically values in the neV to \textmu eV regime can be
reached.  Contrary to backscattering NSE provides an excellent $Q$-resolution.
There exist different methods for neutron spin echo measurements, namely
classical neutron spin echo (NSE) \cite{Mezei:72} and neutron resonance spin
echo (NRSE) \cite{Golub:87(M)}.  The application of NSE and NRSE is currently
limited to measurements with dedicated instruments, where neither the samples
nor the sample environment may depolarise the beam.

A method similar to NRSE is the MIEZE (Modulation of IntEnsity by Zero Effort)
\cite{Gaehler:92} technique.  As in MIEZE all spin manipulations are performed
before the sample position it avoids the complications operating with polarised
neutrons in strong magnetic fields or in a depolarising environment, like
hydrogen containing samples, which introduce spin-flips.  It can easily operate
in any neutron scattering experiment with enough space before the sample and a
detector with nanosecond time resolution.  This method invented by Gähler and
Golub \cite{Gaehler:92} has also been demonstrated experimentally in the
nineties for soft matter samples \cite{Koeppe:96,Hank:97, Besenboeck:98,
  Koeppe:99}.  Recently we published MIEZE measurements on the itinerant magnet
MnSi in a magnetic field \cite{Georgii:11} demonstrating the power of the
method for magnetic applications.

MIEZE is easily implemented as an option at many instruments, provided a fast
detector exists.  It can add energy resolution down to the sub-\textmu eV
region, mainly in the small angle regime.  This paper aims to develop the tools
necessary for designing a MIEZE option for a SANS instrument adding high energy
resolution to it.

\section{Modulation of IntEnsity by Zero Effort (MIEZE)}
\begin{figure}[t]
  \includegraphics[width=\linewidth]{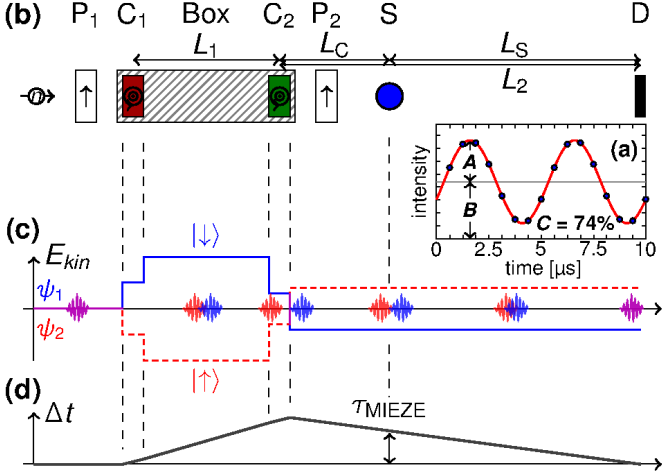}
  \caption{\label{Fig:Schema}{\bf (a)} ~A typical MIEZE signal at the detector
    position (see text for details). {\bf (b)}~Schematic of a complete MIEZE
    setup, showing the polariser (P$_1$), the zero field of the MIEZE box
    (hatched) with two $\pi$-flipper coils (C$_1$, C$_2$), the analyser (P$_2$),
    the sample (S) and the detector (D). {\bf (c)}~Kinetic energy splitting for
    the spin-down ($\psi_1$) and spin-up ($\psi_2$) states of the neutrons along
    the flight path due to the $\pi$-flipper coils. {\bf (d)}~Temporal delay
    $\Delta t$ of the spin states along the flight path.  The splitting reaches
    its maximum after the second flipper coil and vanishes at the detector
    position (after \cite{Georgii:11}).
    \newline
    Reprinted with permission from APS. \copyright 2011, American Institute of Physics.
  }
 \end{figure}

MIEZE uses the first arm of an NRSE instrument, followed by a polarisation
analyser in front of the sample and a fast neutron detector with ns
time-resolution (see figure \ref{Fig:Schema}b).  In the following we use the
wave packet description of the coarse monochromatised neutron beam.  For a
detailed quantum mechanical description see \cite{Arend:04}.  While in NRSE, all
coils are operated at the same frequency \cite{Keller:02} , in MIEZE, the two
coils are driven at different frequencies, $\nu_1=\omega_1/2\pi$ and
$\nu_2=\omega_2/2\pi$, with $\nu_2 > \nu_1$.  This leads to an overcompensation
of the energy splitting of the spin up $\left|\uparrow\right\rangle$ and spin
down $\left|\downarrow\right\rangle$ wave packets in the second coil as shown in
figure \ref{Fig:Schema}c.  As $\left|\uparrow\right\rangle$ and
$\left|\downarrow\right\rangle$ are now propagating with different velocities,
they are interfering only at the specific distance
\begin{equation}
  \label{Ver}
  L_2 = \frac{L_1}{\omega_2 / \omega_1 - 1} 
\end{equation}
after the last coil.  At this position, one obtains a time beating signal
depending on the difference of the two coil frequencies (see figure
\ref{Fig:Schema}a) of the form
\begin{equation}
  I(t) = {\textstyle\frac12} I_0 (1+ C \cdot \cos{\omega_M t}),
\end{equation}
where $\omega_M = 2\cdot(\omega_2 - \omega_1)$ is the frequency difference of
the two NRSE coils.  $C = \frac{I_+ - I_-}{I_+ + I_-}$ is the contrast, which is
given by the ratio of the measured amplitude $A$ and the average intensity $B$
(figure \ref{Fig:Schema}a).

The MIEZE time, which is equivalent to the spin echo time \cite{Keller:02} is
given by
\begin{equation}
  \label{time}
  \tau_M = \frac{\hbar}{mv^3} \omega_M L_S,
\end{equation}
where $L_S$ is the distance between sample and detector.  Similar to
time-of-flight methods, the time resolution obtained depends on $L_S$ as seen in
figure \ref{Fig:Schema}d.

The role of the polarisation in NSE and NRSE is now taken by the contrast $C$ of
the time beating signal, which can be expressed as
\begin{equation}
  C = \int \!\text{d}\omega\, S(\omega) \cos(\omega \tau_M).
\end{equation}
In analogy to NSE, a signal measured at a specific spin-echo time $\tau_M$ is
directly proportional to the intermediate scattering function $S(Q, \tau_M)$.
Thus a typical MIEZE experiment results in the determination of
$S(Q, \tau_M)/S_{el}(Q, \tau_M)$ over $\tau_M$, where $S_{el}(Q,\tau_M)$ is the
signal of an elastic reference sample, usually graphite or the sample in a
frozen state (like very low T in the case of magnetic systems).  For
quasi-elastic experiments with an assumed Lorentzian line shape of half-width
$\Gamma$, the normalised intermediate scattering function is given by
\cite{Keller:02}
\begin{equation}\label{S}
  \frac{S(Q, \tau_M)}{S_{el}(Q, \tau_M)} = \exp\left(-\Gamma(Q)  \tau_M \right).
\end{equation}

\section{Path lengths in MIEZE}
The MIEZE method is closely related to TOF methods and therefore sensitive to
path length differences $\Delta L$.  $\Delta L$ increases for larger $Q$.
Therefore, the two different spin states interfere less with each other leading
to a reduction in the contrast $C$.  The path length differences originate from
different parts of the MIEZE setup and can be expressed by
\def\geom{\text{geometry}}
\begin{multline}
\label{contrast}
  C = R_\text{Coils}
  \cdot R _\text{Sample}(\geom, \Theta, \Lambda) \cdot R_\text{Detector}(\Lambda)  \cdot C_0,
\end{multline}
where $\Theta$ is the scattering angle, $\Lambda = 2\pi v/\omega_M$ the ratio of
neutron velocity and angular frequency of the time-beating signal, i.e. the path
length of one oscillation, and ``$\geom$'' denotes the sample geometry.
$R_\text{Coils}$ contains the contrast reduction in the coil systems in front of
the beam.  $R_\text{Coils}$ is mainly determined by the performance of the
flipper coils and the perfection of the zero field shielding around the system.
$R_\text{Detector}$ treats the loss of contrast due to the thickness of the
detector.  Depending on the interaction depth of the neutrons in the detector,
the flight paths of the neutrons are different, thus reducing the contrast of
the intensity modulation.  As an example, the instrument MIRA \cite{Georgii:07}
at the FRM~II will use a CASCADE detector \cite{Haussler:2011ei,Klein:2011jj}
with 2\,\textmu m thick neutron detection planes. Therefore, $R_\text{Detector}$
is approximately 1.

The reduction factors $R_\text{Coils}$ and $R_\text{Detector}$ depend only
on instrument specific parameters, therefore they can be determined by experiment or
theoretical calculations independently of a specific sample.

In contrast, $R_\text{Sample}$ depends both on the geometry of the experiment
and the sample and needs to be treated separately for each experiment.  While
\citet{Hayashida:2008cd} determined this reduction factor through Monte-Carlo
simulations, we present here analytical formulae, which can be calculated faster
and provide more insight into the influence of different sample geometries on
$C$.  For simplicity we neglect here the influence of the divergence in the
beam, as it anyhow is for SANS quite small.

\begin{figure}
  \begin{center}
    \includegraphics[width=0.8\linewidth]{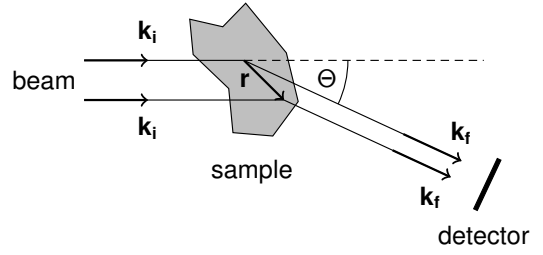}
  \end{center}
  \caption{\label{fig:geometry} The scattering geometry: A parallel beam of
    neutrons is scattered by the whole sample volume under the angle $\Theta$.}
\end{figure}

\begin{figure}
  \begin{center}
    \includegraphics[width=\linewidth]{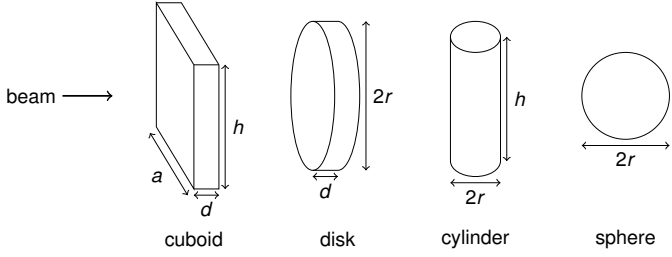}
  \end{center}
  \caption{\label{fig:shapes} Definition of the dimensions of different sample
    shapes.}
\end{figure}

\let\vec=\mathbf

The path length difference $\Delta L$ caused by scattering of a parallel
incoming beam in a sample at two different positions of interaction separated by
$\vec{r}$ (see figure \ref{fig:geometry}) is given in first order by
\cite{Keller:93}
\begin{equation}
  \Delta L(\vec{r}) = \left( \frac{\vec{k_i}}{|\vec{k_i}|} -
                             \frac{\vec{k_f}}{|\vec{k_f}|} \right) \cdot \vec{r}.
\end{equation}
This corresponds to a phase shift of $\Delta \phi(\vec{r}) = \omega_M \Delta
L(\vec{r})/v = 2\pi \Delta L(\vec r)/\Lambda$, where $\Lambda$, as defined
above, is the path length for a single oscillation.  Integrating over the total
sample volume $V$ yields
\begin{eqnarray}
C &= &\int \text{d}\omega\, S(\omega) \frac{1}{V}\int_\text{Sample} \kern -7mm\text{d}^3r\,\cos(\omega \tau_M + \Delta \phi(\vec{r}))\nonumber \\
&=&\int \text{d}\omega\, S(\omega) \frac{1}{V}\int_\text{Sample} \kern -7mm\text{d}^3r\,\big(\cos(\omega \tau_M)  \cos\Delta\phi(\vec{r}) - \nonumber \\
&&\qquad\qquad\qquad\qquad  \sin(\omega \tau_M)  \sin\Delta\phi(\vec{r}) \big) \,.
\end{eqnarray}
For the Lorentzian $S(\omega)$ assumed earlier, the sin-term vanishes when integrating over
$\omega$ and the integration separates into
\begin{equation} 
  C =  \underbrace{\frac{1}{V}\int_\text{Sample} \kern -7mm\text{d}^3r\, \cos\Delta\phi(\vec{r})}_{R_\text{Sample}}
       \underbrace{\int_{\vphantom{\text{Sample}}} \text{d}\omega\, S(\omega) \cos(\omega \tau_M)}_{C_0}.
\end{equation}

From this equation and the geometry given in figure \ref{fig:geometry} we can
derive the correction factor for different sample shapes, with the dimensions
given in figure \ref{fig:shapes}:

\def\thhalf{\frac{\Theta}{2}}

\begin{enumerate}
\item Sphere with radius $r$:
\begin{equation*}
  R(r, \Theta, \Lambda) = \frac{3\Lambda^3}{64\pi^3r^3}
  \frac{\sin\left(\frac{4\pi r}{\Lambda} \sin\thhalf\right)}{\sin^3\thhalf} 
- \frac{3\Lambda^2 \cos\left(\frac{4\pi r}{\Lambda} \sin\thhalf\right)}{16\pi^2r^2}.
\end{equation*}

\item Cylinder with radius $r$:
\begin{equation*}
  R(r, \Theta, \Lambda) = \frac{\Lambda}{2\pi r}
    \frac{J_1\left(\frac{4\pi r}{\Lambda} \sin\thhalf\right)}{\sin\thhalf},
\end{equation*}
where $J_1$ is the Bessel function of the first kind.  The height $h$ of the
cylinder -- as it is oriented perpendicular to the scattering plane -- does not
affect the reduction factor $R$ for a parallel incoming beam.

\item Cuboid with thickness $d$ and width $a$:
\begin{equation*}
  R(d,a,\Theta,\Lambda) = \frac{\Lambda^2}{4\pi^2da}
  \frac{\sin\left(\frac{2\pi a}{\Lambda}\cos\thhalf\sin\thhalf\right)
        \sin\left(\frac{2\pi d}{\Lambda}\sin^2\thhalf\right)} {\sin^3\thhalf\cos\thhalf}.
\end{equation*}
As for the cylinder the reduction factor R does not depend on $h$.

\item Disk with thickness $d$ and radius $r$:
\begin{eqnarray*}
  R(d,r,\Theta,\Lambda)&=&\frac{\Lambda^2}{4\pi^3 d r^2}
  \frac{\sin\left(\frac{2\pi d}{\Lambda}\sin^2\thhalf\right)}{\sin^3\thhalf\cos\thhalf} \cdot\\
  &&\int_{-r}^{r}\!\!\text{d}z\,\sin\left(\frac{2\pi\sqrt{r^2-z^2}}{\Lambda}\sin\Theta\right).
\end{eqnarray*}

\end{enumerate}

\begin{figure}[t]
  \begin{center}
   \includegraphics[width=\linewidth]{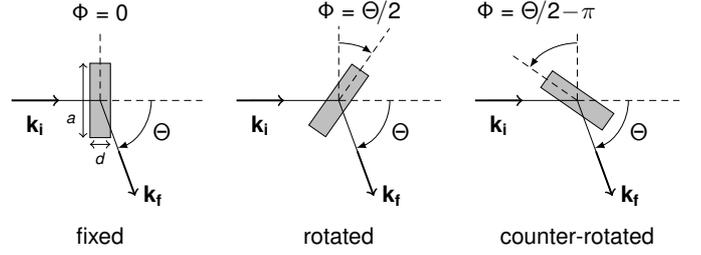}\\[1ex]
   \includegraphics[width=0.9\linewidth]{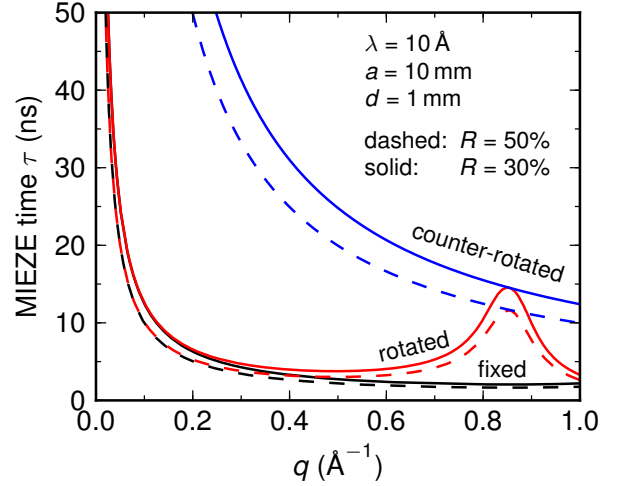}
 \end{center} \caption{\label{Cuboid_rot} The $Q$, $\tau_M$ parameter space for a
   MIEZE instrument with $\lambda=10$\AA, and a cuboid sample with $a=10$\,mm
   and $d=1$\,mm.  The lines indicate where the contrast
   for different sample configurations is reduced to 50\% and 30\%.  The sample
   is either not rotated at all, or rotated by half the scattering angle with
   respect to two different positions at $\Theta=0$.}
\end{figure}

By rotation of any plate-like sample (cuboid or disk) by $\Phi=\Theta/2-\pi$
(called ``counter-rotation'', see figure \ref{Cuboid_rot}), one obtains:
\begin{eqnarray*}
  R(d,\Theta,\Lambda) &= &\frac{\Lambda}{2\pi d}
  \frac{\sin\left(\frac{2\pi d}{\Lambda}\sin\thhalf\right)}{\sin\thhalf}.
\end{eqnarray*}
This results in a much slower decrease of $R$ with increasing $\Theta$ as $R$
depends only on $d$, the thickness of the sample, which can be made small.

\begin{figure}
  \begin{center}
   \includegraphics[width=0.9\linewidth]{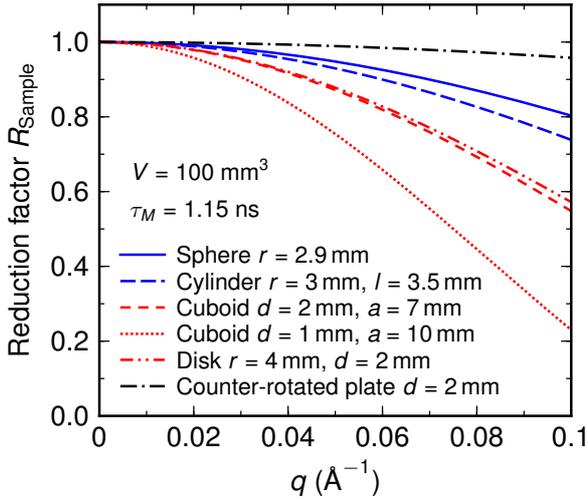}
  \end{center}
  \caption{\label{Geometries} The reduction factor $R_\text{Sample}$ for different
   geometries with the same sample volume of $100\, \text{mm}^3$.  It is
   calculated for the instrument parameters of MIRA with a wavelength $\lambda$
   of 10.4\,\AA.  $L_1$ is 1\,m, and $L_S$ is 0.8\,m.  $\omega_M$ is $2 \pi
   \cdot 200$ kHz, corresponding to $\tau_M=1.15$\,ns.  The scattering geometry
   is defined as shown in figure \ref{fig:geometry}.}
\end{figure}

In figure \ref{Geometries} the reduction factor $R_\text{Sample}$ is plotted
versus $Q = \frac{4\pi}{\lambda} \sin{\thhalf}$ for different geometries of the
sample.  It becomes obvious that differences are only important for larger $Q$
values, and that there are large differences between different sample shapes.

In figure \ref{Cuboid_rot} the effect of sample rotation for measurements on
cuboidal samples is demonstrated.  The accessible parameter space in $Q$ and
$\tau_M$ can be enlarged when turning the sample by half the scattering
angle, in the right direction.

\begin{figure}[t]
  \begin{center}
   \includegraphics[width=0.9\linewidth]{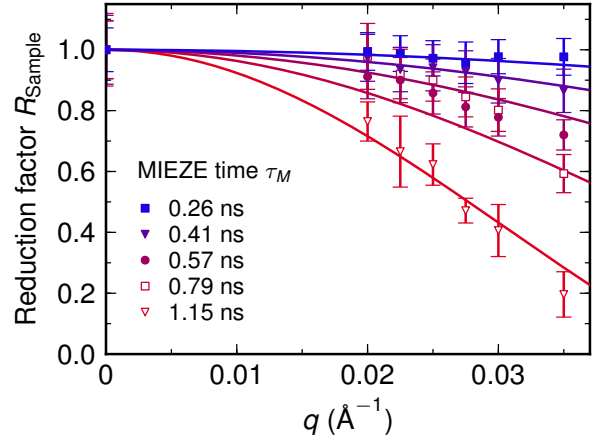}
  \end{center} \caption{\label{graphite} The reduction factor $R$ versus $Q$ for a
  cuboid of width $a=25$\,mm and thickness $d=5$\,mm compared to measured values
  of $R$ on the instrument MIRA with a wavelength $\lambda$ of 10.4\,\AA. $L_1$
  is 1\,m, and $L_S$ is 0.8\,m.  $\omega_M$ ranges from $2 \pi \cdot 46$\,kHz to
  $2 \pi \cdot 200$\,kHz, yielding a MIEZE time $\tau_M$ from 0.26\,ns to
  1.15\,ns, respectively.}
\end{figure}

These theoretical predictions were tested for a cuboid of thickness $d=5$\,mm
and width $a=25$\,mm using the MIEZE setup at FRM II \cite{Georgii:11} and a
very good agreement for various $\tau_M$ is obtained as shown in figure
\ref{graphite}.

\section{High resolution MISANS at ESS}

From the discussion above it becomes clear that the MIEZE technique is
particularly well suited for measurements in the small angle regime as for low
$Q$ the contrast reduction due to the path length differences is less severe.
Therefore a combination of a time-of-flight SANS instrument with a MIEZE option
(MISANS) would allow for high resolution measurements both in energy- and
Q-space.  The MIEZE principle at a time-of-flight source was recently
demonstrated at the chopped CG-1D beam at HFIR at the Oak Ridge National
Laboratory \cite{Brandl:2011}.

Equations (\ref{Ver}) and (\ref{time}) demonstrate that the different instrument
design parameters are correlated.  If $L_2$ is replaced by $L_2 = L_C +L_S$,
where $L_C$ is the distance between the last coil and the sample (see figure
\ref{Fig:Schema}b), one obtains
\begin{equation}
  \label{eq:taum}
  \tau_M = \frac{2  \hbar \omega_1 L_S L_1}{m v^3 (L_S + L_C)}\,.
\end{equation}
This equation now allows for trading off different instrument designs if one
defines the maximum range of Spin-Echo times.

Current NSE measurements are performed up to 1\,\textmu s \cite{Ohl:04} and are
a benchmark for new spin echo beam lines.  Considering typical SANS setups as
proposed for long-pulse spallation sources such as the ESS \cite{Schober:2008} a
zero-field region can be added to the 20\,m long collimation section, with a
coil distance, for example, $L_1 = 15$\,m, a coil-sample distance $L_C=5$\,m and
a sample-detector distance $L_S=10$\,m.  For neutrons with $\lambda=20$\,\AA,
which is a typical wavelength used in NSE for high resolution measurements, the
remaining free parameter in eq. (\ref{eq:taum}) is $\omega_1$.  To achieve
$\tau_M = 1$\,\textmu s with this setup, the coils have to be driven at
$\omega_1 = 2\pi\cdot 1$\,MHz and $\omega_2 = 2\pi\cdot 2$\,MHz so that
$\omega_M=2\pi\cdot 2$\,MHz.  This can be obtained with current coil designs: RF
coils that can be driven at these frequencies are in commissioning at the
instruments RESEDA \cite{Haus:08} and TRISP \cite{Keller:07} at FRM~II.

\begin{figure}[tbh]
  \begin{center}
  \includegraphics[width=0.9\linewidth]{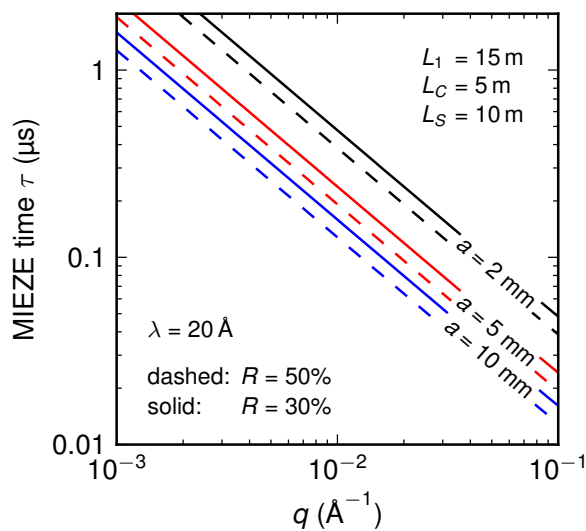}
  \end{center}
  \caption{\label{MISANS} The $Q$, $\tau_M$ parameter space for a MISANS
    instrument at the ESS, assuming a fixed cuboid sample of several different
widths $a$ and a thickness of $b=2$\,mm.  It is shown where the contrast for the
different samples is reduced to 50\% (the dashed lines), or to 30\% (the solid
lines).}
\end{figure}

For these instrument parameters, a $\tau_M$--$Q$ parameter space is opened as
shown in figure \ref{MISANS} for cuboid samples of several sizes.  MIEZE times
of 1\,\textmu s are achieved up to $Q=5\times10^{-3}$\,\AA$^{-1}$ for samples
with a width of 5\,mm and a thickness of 2\,mm.  The largest spin echo times are
available for small $Q$, which matches the requirements of measuring
quasi-elastic dynamics, i.e. very slow relaxation processes as they are expected
for large scale structures, for example in soft matter and magnetic materials
with novel topological structures.

We do not discuss the $Q$ resolution of such an instrument here, as it is mainly
defined by $\Delta\lambda/\lambda$, which is already small at a pulsed source,
$\approx 3 \%$ at the ESS as discussed in \cite{Schober:2008}, and by the beam
divergence, which will also be excellent for such a long collimation section.

In conclusion, we propose to rather dramatically enhance the capabilities of a
SANS beam line at the ESS with different options to obtain in parallel
information on structure {\it and} dynamics, especially in magnetic fields.
These options can be MIEZE, TISANE \cite{2008PhLA..372.1541K} and stroboscopic
SANS \cite{Keiderling:2007jc}.  The latter two are basically available free of
charge with the fast time-resolving detector for MIEZE.  They will cover the
time domain from nanoseconds to minutes.  Such an instrument based on existing
technology would open new perspectives for research in magnetic systems and soft
matter.  It would offer an excellent $q$-resolution and at the same time allow
to measure the dynamics on a wide range of time scales\footnote{Strictly
  speaking, stroboscopic SANS and TISANE are only able to
  resolve dynamics stimulated by a periodic signal, whereas MIEZE has the
  potential to observe the dynamics in thermal equilibrium through the
  interaction of the neutrons with the system.}, competitive to NSE or NRSE
instruments.  It would also be much cheaper and easier to build due to the
reduced effort in magnetic shielding. Furthermore it allows to use modern
focusing neutron optics in any part of the instrument, enhancing the intensity
at the sample. 

\section{Acknowledgment}
We acknowledge very helpful discussions with J.~Neuhaus, T.~Keller, K.~Habicht, C.~Pfleiderer, M.~Bleuel and J.~Lal.
This work is supported by the BMBF under ``Mitwirkung der Zentren der Helmholtz
Gemeinschaft und der Technischen Uni\-ver\-sit\"at M\"un\-chen an der Design-Update
Phase der ESS, F\"or\-der\-kenn\-zeichen 05E10WO1.''


\begin{thebibliography}{24}
\providecommand{\natexlab}[1]{#1}
\providecommand{\url}[1]{\texttt{#1}}
\providecommand{\urlprefix}{URL }
\expandafter\ifx\csname urlstyle\endcsname\relax
  \providecommand{\doi}[1]{doi:\discretionary{}{}{}#1}\else
  \providecommand{\doi}[1]{doi:\discretionary{}{}{}\begingroup
  \urlstyle{rm}\url{#1}\endgroup}\fi
\providecommand{\bibinfo}[2]{#2}

\bibitem[{Mezei(1972)}]{Mezei:72}
\bibinfo{author}{F.~Mezei}, \bibinfo{title}{Neutron spin echo: A new concept in
  polarized thermal neutron techniques}, \bibinfo{journal}{Zeitschrift für
  Physik A} \bibinfo{volume}{255}~(\bibinfo{number}{2}) (\bibinfo{year}{1972})
  \bibinfo{pages}{146--160}, \doi{\bibinfo{doi}{10.1007/BF01394523}}.

\bibitem[{Richter et~al.(2005)Richter, Monkenbusch, Arbe, and
  Colmenero}]{Richter:05}
\bibinfo{author}{D.~Richter}, \bibinfo{author}{M.~Monkenbusch},
  \bibinfo{author}{A.~Arbe}, \bibinfo{author}{J.~Colmenero},
  \bibinfo{title}{Neutron Spin Echo in Polymer Systems},
  \bibinfo{journal}{Advances in Polymer Science} \bibinfo{volume}{174}
  (\bibinfo{year}{2005}) \bibinfo{pages}{1--221},
  \doi{\bibinfo{doi}{10.1007/b106578}}.

\bibitem[{Mezei(1984)}]{Mezei:84}
\bibinfo{author}{F.~Mezei}, \bibinfo{title}{Critical Dynamics In Isotropic
  Ferromagnets}, \bibinfo{journal}{Journal of Magnetism and Magnetic Materials}
  \bibinfo{volume}{45}~(\bibinfo{number}{1}) (\bibinfo{year}{1984})
  \bibinfo{pages}{67--73}, \doi{\bibinfo{doi}{10.1016/0304-8853(84)90374-3}}.

\bibitem[{Golub and Gähler(1987)}]{Golub:87(M)}
\bibinfo{author}{R.~Golub}, \bibinfo{author}{R.~Gähler}, \bibinfo{title}{A
  neutron resonance spin echo spectrometer for quasi-elastic and inelastic
  scattering}, \bibinfo{journal}{Physics Letters A}
  \bibinfo{volume}{123}~(\bibinfo{number}{1}) (\bibinfo{year}{1987})
  \bibinfo{pages}{43--48}, ISSN \bibinfo{issn}{0375-9601},
  \doi{\bibinfo{doi}{10.1016/0375-9601(87)90760-2}}.

\bibitem[{Gähler et~al.(1992)Gähler, Golub, and Keller}]{Gaehler:92}
\bibinfo{author}{R.~Gähler}, \bibinfo{author}{R.~Golub},
  \bibinfo{author}{T.~Keller}, \bibinfo{title}{Neutron resonance spin echo--a
  new tool for high resolution spectroscopy}, \bibinfo{journal}{Physica B:
  Condensed Matter} \bibinfo{volume}{180-181}~(\bibinfo{number}{2})
  (\bibinfo{year}{1992}) \bibinfo{pages}{899--902},
  \doi{\bibinfo{doi}{10.1016/0921-4526(92)90503-K}}.

\bibitem[{Köppe et~al.(1996)Köppe, Hank, Wuttke, Petry, Gähler, and
  Kahn}]{Koeppe:96}
\bibinfo{author}{M.~Köppe}, \bibinfo{author}{P.~Hank},
  \bibinfo{author}{J.~Wuttke}, \bibinfo{author}{W.~Petry},
  \bibinfo{author}{R.~Gähler}, \bibinfo{author}{R.~Kahn},
  \bibinfo{title}{Performance and future of a neutron resonance spinecho
  spectrometer}, \bibinfo{journal}{Journal for Neutron Research}
  \bibinfo{volume}{4}~(\bibinfo{number}{1}) (\bibinfo{year}{1996})
  \bibinfo{pages}{261--273}, \doi{\bibinfo{doi}{10.1080/10238169608200092}}.

\bibitem[{Hank et~al.(1997)Hank, Besenböck, Gähler, and Köppe}]{Hank:97}
\bibinfo{author}{P.~Hank}, \bibinfo{author}{W.~Besenböck},
  \bibinfo{author}{R.~Gähler}, \bibinfo{author}{M.~Köppe},
  \bibinfo{title}{Zero-field neutron spin echo techniques for incoherent
  scattering}, \bibinfo{journal}{Physica B: Condensed Matter}
  \bibinfo{volume}{234-236} (\bibinfo{year}{1997}) \bibinfo{pages}{1130--1132},
  \doi{\bibinfo{doi}{10.1016/S0921-4526(97)89269-1}}.

\bibitem[{Besenböck et~al.(1998)Besenböck, Gähler, Hank, Kahn, Köppe,
  de~Novion, Petry, and Wuttke}]{Besenboeck:98}
\bibinfo{author}{W.~Besenböck}, \bibinfo{author}{R.~Gähler},
  \bibinfo{author}{P.~Hank}, \bibinfo{author}{R.~Kahn},
  \bibinfo{author}{M.~Köppe}, \bibinfo{author}{C.-H. de~Novion},
  \bibinfo{author}{W.~Petry}, \bibinfo{author}{J.~Wuttke},
  \bibinfo{title}{First scattering experiment on MIEZE: A fourier transform
  time-of-flight spectrometer using resonance coils}, \bibinfo{journal}{Journal
  for Neutron Research} \bibinfo{volume}{7}~(\bibinfo{number}{1})
  (\bibinfo{year}{1998}) \bibinfo{pages}{65--74},
  \doi{\bibinfo{doi}{10.1080/10238169808200231}}.

\bibitem[{Köppe et~al.(1999)Köppe, Bleuel, Gähler, Golub, Hank, Keller,
  Longeville, Rauch, and Wuttke}]{Koeppe:99}
\bibinfo{author}{M.~Köppe}, \bibinfo{author}{M.~Bleuel},
  \bibinfo{author}{R.~Gähler}, \bibinfo{author}{R.~Golub},
  \bibinfo{author}{P.~Hank}, \bibinfo{author}{T.~Keller},
  \bibinfo{author}{S.~Longeville}, \bibinfo{author}{U.~Rauch},
  \bibinfo{author}{J.~Wuttke}, \bibinfo{title}{Prospects of resonance spin
  echo}, \bibinfo{journal}{Physica B: Condensed Matter}
  \bibinfo{volume}{266}~(\bibinfo{number}{1-2}) (\bibinfo{year}{1999})
  \bibinfo{pages}{75--86}, \doi{\bibinfo{doi}{10.1016/S0921-4526(98)01496-3}}.

\bibitem[{Georgii et~al.(2011)Georgii, Brandl, Arend, Häußler, Tischendorf,
  Pfleiderer, Böni, and Lal}]{Georgii:11}
\bibinfo{author}{R.~Georgii}, \bibinfo{author}{G.~Brandl},
  \bibinfo{author}{N.~Arend}, \bibinfo{author}{W.~Häußler},
  \bibinfo{author}{A.~Tischendorf}, \bibinfo{author}{C.~Pfleiderer},
  \bibinfo{author}{P.~Böni}, \bibinfo{author}{J.~Lal},
  \bibinfo{title}{Turn-key module for neutron scattering with sub-micro-eV
  resolution}, \bibinfo{journal}{Applied Physics Letters}
  \bibinfo{volume}{98}~(\bibinfo{number}{7}) (\bibinfo{year}{2011})
  \bibinfo{pages}{073505}, \doi{\bibinfo{doi}{10.1063/1.3556558}}.

\bibitem[{Arend et~al.(2004)Arend, Gähler, Keller, Georgii, Hils, and
  Böni}]{Arend:04}
\bibinfo{author}{N.~Arend}, \bibinfo{author}{R.~Gähler},
  \bibinfo{author}{T.~Keller}, \bibinfo{author}{R.~Georgii},
  \bibinfo{author}{T.~Hils}, \bibinfo{author}{P.~Böni},
  \bibinfo{title}{Classical and quantum-mechanical picture of NRSE--measuring
  the longitudinal Stern-Gerlach effect by means of TOF methods},
  \bibinfo{journal}{Physics Letters A}
  \bibinfo{volume}{327}~(\bibinfo{number}{1}) (\bibinfo{year}{2004})
  \bibinfo{pages}{21--27}, \doi{\bibinfo{doi}{10.1016/j.physleta.2004.04.062}}.

\bibitem[{Keller et~al.(2002)Keller, Golub, and Gähler}]{Keller:02}
\bibinfo{author}{T.~Keller}, \bibinfo{author}{R.~Golub},
  \bibinfo{author}{R.~Gähler}, \bibinfo{title}{Neutron Spin Echo--A Technique
  for High-Resolution Neutron Scattering}, in: \bibinfo{editor}{R.~Pike},
  \bibinfo{editor}{P.~Sabatier} (Eds.), \bibinfo{booktitle}{Scattering},
  \bibinfo{publisher}{Academic Press}, \bibinfo{address}{London}, ISBN
  \bibinfo{isbn}{978-0-12-613760-6}, \bibinfo{pages}{1264--1286},
  \doi{\bibinfo{doi}{10.1016/B978-012613760-6/50068-1}}, \bibinfo{year}{2002}.

\bibitem[{Georgii et~al.(2007)Georgii, B{\"o}ni, Janoschek, Schanzer, and
  Valloppilly}]{Georgii:07}
\bibinfo{author}{R.~Georgii}, \bibinfo{author}{P.~B{\"o}ni},
  \bibinfo{author}{M.~Janoschek}, \bibinfo{author}{C.~Schanzer},
  \bibinfo{author}{S.~Valloppilly}, \bibinfo{title}{MIRA--A flexible instrument
  for VCN}, \bibinfo{journal}{Physica B}
  \bibinfo{volume}{397}~(\bibinfo{number}{1-2}) (\bibinfo{year}{2007})
  \bibinfo{pages}{150--152}, \doi{\bibinfo{doi}{10.1016/j.physb.2007.02.088}}.

\bibitem[{Häußler et~al.(2011)Häußler, Böni, Klein, Schmidt, Schmidt,
  Groitl, and Kindervater}]{Haussler:2011ei}
\bibinfo{author}{W.~Häußler}, \bibinfo{author}{P.~Böni},
  \bibinfo{author}{M.~Klein}, \bibinfo{author}{C.~J. Schmidt},
  \bibinfo{author}{U.~Schmidt}, \bibinfo{author}{F.~Groitl},
  \bibinfo{author}{J.~Kindervater}, \bibinfo{title}{Detection of high frequency
  intensity oscillations at RESEDA using the CASCADE detector},
  \bibinfo{journal}{Review of Scientific Instruments}
  \bibinfo{volume}{82}~(\bibinfo{number}{4}) (\bibinfo{year}{2011})
  \bibinfo{pages}{045101}, \doi{\bibinfo{doi}{10.1063/1.3571300}}.

\bibitem[{Klein and Schmidt(2011)}]{Klein:2011jj}
\bibinfo{author}{M.~Klein}, \bibinfo{author}{C.~J. Schmidt},
  \bibinfo{title}{CASCADE, neutron detectors for highest count rates in
  combination with ASIC/FPGA based readout electronics}, in:
  \bibinfo{booktitle}{Nuclear Instruments and Methods in Physics Research
  Section A: Accelerators, Spectrometers, Detectors and Associated Equipment},
  vol. \bibinfo{volume}{628}, \bibinfo{pages}{9--18},
  \doi{\bibinfo{doi}{10.1016/j.nima.2010.06.278}}, \bibinfo{year}{2011}.

\bibitem[{Hayashida et~al.(2008)Hayashida, Hino, Kitaguchi, Kawabata, and
  Achiwa}]{Hayashida:2008cd}
\bibinfo{author}{H.~Hayashida}, \bibinfo{author}{M.~Hino},
  \bibinfo{author}{M.~Kitaguchi}, \bibinfo{author}{Y.~Kawabata},
  \bibinfo{author}{N.~Achiwa}, \bibinfo{title}{A study of resolution function
  on a MIEZE spectrometer}, \bibinfo{journal}{Measurement Science and
  Technology} \bibinfo{volume}{19} (\bibinfo{year}{2008})
  \bibinfo{pages}{4006}, \doi{\bibinfo{doi}{10.1088/0957-0233/19/3/034006}}.

\bibitem[{Keller(1993)}]{Keller:93}
\bibinfo{author}{T.~Keller}, \bibinfo{title}{Höchstauflösende
  Neutronenspektrometer auf Basis von Spinflippern -- neue Varianten des
  Spinecho-Prinzips}, Ph.D. thesis, \bibinfo{school}{Technische Universität
  München}, \bibinfo{year}{1993}.

\bibitem[{Brandl et~al.(2011)Brandl, Georgii, Bleuel, Carpenter, Robertson,
  Crow, and Lal}]{Brandl:2011}
\bibinfo{author}{G.~Brandl}, \bibinfo{author}{R.~Georgii},
  \bibinfo{author}{M.~Bleuel}, \bibinfo{author}{J.~Carpenter},
  \bibinfo{author}{L.~Robertson}, \bibinfo{author}{L.~Crow},
  \bibinfo{author}{J.~Lal}, \bibinfo{title}{Measurements on MnSi with MISANS},
  \bibinfo{journal}{to be published in: Journal of Physics: Conference Series}
  .

\bibitem[{Ohl et~al.(2004)Ohl, Monkenbusch, Richter, and Pappas}]{Ohl:04}
\bibinfo{author}{M.~Ohl}, \bibinfo{author}{M.~Monkenbusch},
  \bibinfo{author}{D.~Richter}, \bibinfo{author}{C.~Pappas},
  \bibinfo{title}{The high-resolution neutron spin-echo spectrometer for the
  SNS with $\tau < \mu$s}, \bibinfo{journal}{Physica B: Condensed Matter}
  \bibinfo{volume}{350} (\bibinfo{year}{2004}) \bibinfo{pages}{147–150},
  \doi{\bibinfo{doi}{10.1016/j.physb.2004.04.014}}.

\bibitem[{Schober et~al.(2008)Schober, Farhi, Mezei, Allenspach, Andersen,
  Bentley, Christiansen, Cubitt, Heenan, Kulda, Langan, Lefmann, Lieutenant,
  Monkenbusch, Willendrup, Saroun, Tindemans, and Zsigmond}]{Schober:2008}
\bibinfo{author}{H.~Schober}, \bibinfo{author}{E.~Farhi},
  \bibinfo{author}{F.~Mezei}, \bibinfo{author}{P.~Allenspach},
  \bibinfo{author}{K.~Andersen}, \bibinfo{author}{P.~M. Bentley},
  \bibinfo{author}{P.~Christiansen}, \bibinfo{author}{B.~Cubitt},
  \bibinfo{author}{R.~K. Heenan}, \bibinfo{author}{J.~Kulda},
  \bibinfo{author}{P.~Langan}, \bibinfo{author}{K.~Lefmann},
  \bibinfo{author}{K.~Lieutenant}, \bibinfo{author}{M.~Monkenbusch},
  \bibinfo{author}{P.~Willendrup}, \bibinfo{author}{J.~Saroun},
  \bibinfo{author}{P.~Tindemans}, \bibinfo{author}{G.~Zsigmond},
  \bibinfo{title}{Tailored instrumentation for long-pulse neutron spallation
  sources}, \bibinfo{journal}{Nuclear Instruments and Methods in Physics
  Research Section A: Accelerators, Spectrometers, Detectors and Associated
  Equipment} \bibinfo{volume}{589}~(\bibinfo{number}{1}) (\bibinfo{year}{2008})
  \bibinfo{pages}{34--46}, \doi{\bibinfo{doi}{10.1016/j.nima.2008.01.102}}.

\bibitem[{Häußler et~al.(2008)Häußler, Streibl, and Böni}]{Haus:08}
\bibinfo{author}{W.~Häußler}, \bibinfo{author}{D.~Streibl},
  \bibinfo{author}{P.~Böni}, \bibinfo{title}{RESEDA: Double and Multi Detector
  Arms for Neutron Resonance Spin Echo Spectrometers},
  \bibinfo{journal}{Measurement Science and Technology} \bibinfo{volume}{19}
  (\bibinfo{year}{2008}) \bibinfo{pages}{034015},
  \doi{\bibinfo{doi}{10.1088/0957-0233/19/3/034015}}.

\bibitem[{Keller et~al.(2007)Keller, Aynajian, Bayrakci, Buchner, Habicht,
  Klann, Ohl, and Keimer}]{Keller:07}
\bibinfo{author}{T.~Keller}, \bibinfo{author}{P.~Aynajian},
  \bibinfo{author}{S.~Bayrakci}, \bibinfo{author}{K.~Buchner},
  \bibinfo{author}{K.~Habicht}, \bibinfo{author}{H.~Klann},
  \bibinfo{author}{M.~Ohl}, \bibinfo{author}{B.~Keimer}, \bibinfo{title}{The
  Triple Axis Spin-Echo Spectrometer TRISP at the FRM II},
  \bibinfo{journal}{Neutron News} \bibinfo{volume}{18}~(\bibinfo{number}{2})
  (\bibinfo{year}{2007}) \bibinfo{pages}{16--18},
  \doi{\bibinfo{doi}{10.1080/10448630701328372}}.

\bibitem[{Kipping et~al.(2008)Kipping, G{\"a}hler, and
  Habicht}]{2008PhLA..372.1541K}
\bibinfo{author}{D.~Kipping}, \bibinfo{author}{R.~G{\"a}hler},
  \bibinfo{author}{K.~Habicht}, \bibinfo{title}{Small angle neutron scattering
  at very high time resolution: Principle and simulations of 'TISANE'},
  \bibinfo{journal}{Physics Letters A}
  \bibinfo{volume}{372}~(\bibinfo{number}{10}) (\bibinfo{year}{2008})
  \bibinfo{pages}{1541--1546},
  \doi{\bibinfo{doi}{10.1016/j.physleta.2007.10.041}}.

\bibitem[{Keiderling and Wiedenmann(2007)}]{Keiderling:2007jc}
\bibinfo{author}{U.~Keiderling}, \bibinfo{author}{A.~Wiedenmann},
  \bibinfo{title}{Field-dependent relaxation behavior of Co-ferrofluid
  investigated with stroboscopic time-resolved small-angle neutron scattering},
  \bibinfo{journal}{Journal of Applied Crystallography}
  \bibinfo{volume}{40}~(\bibinfo{number}{s1}) (\bibinfo{year}{2007})
  \bibinfo{pages}{s62--s67}, \doi{\bibinfo{doi}{10.1107/S0021889806047650}}.

\end{thebibliography}

\end{document}